\documentclass[useAMS,usenatbib]{mn2e}
\usepackage{graphicx}
\usepackage{amssymb}
\usepackage{amsmath}
\usepackage{verbatim}
\usepackage{booktabs}
\usepackage{mathtools}
\usepackage{cuted}
\usepackage[export]{adjustbox}
\usepackage{url}
\usepackage{calrsfs}
\newcommand{\wrot}{\omega_{\text{rot}}}

\newcommand{\worb}{\omega_{\text{orb}}}

\newcommand{\thobs}{\theta_{o}}
\newcommand{\phobs}{\phi_{o}}
\newcommand{\thst}{\theta_{s}}
\newcommand{\phst}{\phi_{s}}
\newcommand{\xis}{\xi_{\text{s}}}
\newcommand{\phobsz}{\phi_{o}(0)}

\title[Analytic Reflected Lightcurves]{Analytic Reflected Lightcurves for Exoplanets}
\author[H.M.~Haggard and N.B.~Cowan]{Hal~M.~Haggard$^1$ and Nicolas~B.~Cowan$^{2}$\\
$^1$ Physics Program, Bard College, 30 Campus Road, Annondale-On-Hudson, NY 12504, USA,\\
Perimeter Institute for Theoretical Physics, 31 Caroline Street North, Waterloo, ON, N2L 2Y5, CAN\\
$^2$ Department of Earth \& Planetary Sciences, McGill University, 3450 rue University, Montreal, QC, H3A 0E8, CAN, \\ Department of Physics, McGill University, 3600 rue University, Montreal, QC, H3A 2T8, CAN,\\
McGill Space Institute, 3550 rue University, Montreal, QC, H3A 2A7, CAN, and\\
Institut de recherche sur les exoplan\`etes, Universit\'e de Montr\'eal, C.P. 6128, Succ.\ Centre-ville, Montr\'eal QC H3C 3J7, CAN}

\begin{document}

\maketitle

\begin{abstract}
The disk-integrated reflected brightness of an exoplanet changes as a function of time due to orbital and rotational motion coupled with an inhomogeneous albedo map. 
We have previously derived analytic reflected lightcurves for spherical harmonic albedo maps in the special case of a synchronously-rotating planet on an edge-on orbit (Cowan, Fuentes \& Haggard 2013).  In this paper, we present analytic reflected lightcurves for the general case of a planet on an inclined orbit, with arbitrary spin period and non-zero obliquity. We do so for two different albedo basis maps: bright points ($\delta$-maps), and spherical harmonics ($Y_l^m$-maps). 
In particular, we use Wigner $D$-matrices to express an harmonic lightcurve for an arbitrary viewing geometry as a non-linear combination of harmonic lightcurves for the simpler edge-on, synchronously rotating geometry. 
These solutions will enable future exploration of the degeneracies and information content of reflected lightcurves, as well as fast calculation of lightcurves for mapping exoplanets based on time-resolved photometry. To these ends we make available Exoplanet Analytic Reflected Lightcurves (EARL), a simple open-source code that allows rapid computation of reflected lightcurves.          
\end{abstract}

%Post to astro-ph.EP and astro-ph.IM

 \section{Introduction}
We would like to know what exoplanets look like, but they are too distant to resolve with near-term telescopes. Fortunately, the rotational and orbital motion of exoplanets bring different features in and out of view---the resulting changes in brightness and color betray the planet's appearance (for a recent review of exo-cartography, see Cowan \& Fujii 2017). An exoplanet's reflected light contribution to total system brightness is of order $10^{-5}$ for a hot Jupiter, and closer to $10^{-7}$ for a hot Earth. Measuring these small contributions to the system flux is currently feasible with the \emph{Kepler} telescope. For an Earth-like planet orbiting a Sun-like star, the contrast is of order $10^{-9}$, which will have to wait until future missions like HabEx and LUVOIR.  

Even once we are able to measure the brightness variations due to reflected light from a planet, it is non-trivial to invert these data to obtain an albedo map of the planet. \cite{Ford_2001} demonstrated that the brightness and color variations of the proverbial ``pale blue dot'' encodes information about surface and atmospheric features on the planet.  But the inverse problem---inferring planetary properties from disk-integrated photometry---has proven a much tougher problem. It took over a decade to solve isolated aspects of the problem: establishing rotation frequency \citep{Palle_2008, Oakley_2009}, rotational (1-D) mapping \citep{Cowan_2009, Cowan_2011,Fujii_2011}, spin-orbit (2-D) mapping \citep{Kawahara_2010,Kawahara_2011,Fujii_2012}, and retrieval of individual surface spectra \citep{Fujii_2010,Cowan_Strait_2013,Fujii_2017}.

In parallel, there have been theoretical and observational efforts to map the atmospheres of short-period exoplanets, both in reflected light and thermal emission, the latter being observationally and mathematically easier \citep{Williams_2006, Rauscher_2007,Knutson_2007a,Knutson_2012,Cowan_Agol_2008,Majeau_2012,deWit_2012,Cowan_2013,Demory_2013,Cowan_2017}.  The primary differences between the direct imaging and combined light approaches are that short-period planets are expected to have zero obliquity and are probably synchronously rotating, which makes the geometry simpler \citep[but see][]{Leconte_2015, Dang_2018}. In this paper, we focus on the hardest version of the problem: reflected light from a planet with arbitrary orbital inclination and spin.   

\begin{figure*}%[htb]%  figure placement: here, top, bottom, or page
   \begin{center}
   \vspace{-4.0cm}
  %\hspace{+1.0cm}
  \includegraphics[width=1.10\textwidth, center]{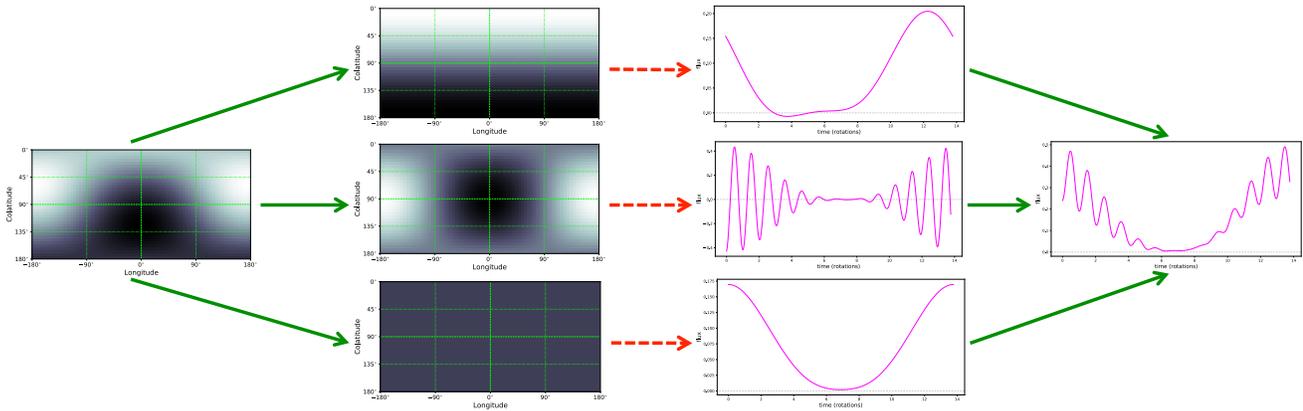} %\hspace{-2.0cm} 
   \vspace{-5cm}
   \caption{An arbitrary planetary albedo map (far left) can be decomposed into spherical harmonics (left), each of which has a geometry-dependent harmonic lightcurve (right).  The lightcurve of the arbitrary map is simply the sum of the harmonic lightcurves, weighted by the coefficients of the spherical harmonics (far right). The colors of arrows denote the difficulty of the operation: decomposing a map into spherical harmonics is simply  integration, and the operation is unnecessary if the map is parameterized using spherical harmonics to begin with. Combining weighted harmonic lightcurves to obtain the overall lightcurve is a simple matter of addition. The crux of the forward problem is computing the harmonic lightcurves for a given viewing geometry (red, dashed arrows).  The brute force approach is numerical integration, but in this paper we analytically solve the map-to-lightcurve transformation. (Insets produced using ReflectDirect, created by J.C.~Schwartz.)} 
   \label{flow_chart}
\end{center}
\end{figure*}

In order to solve the inverse problem of exo-cartography, it is helpful---if not necessary---to have a firm grasp of the forward problem: calculating the brightness variations of a planet given its surface inhomogeneities, as well as orbital and viewing geometry.  A particularly useful approach to the forward problem is to decompose the map using a set of orthonormal basis maps for which one can analytically determine the resulting lightcurves (Figure~\ref{flow_chart}). Doing this analytically makes it easier to track down degeneracies and special cases.  Moreover, it is necessary to adopt a parameterized map when solving the inverse problem, and the simplest way to do so is to have each of the map parameters correspond to a single orthonormal-basis-component map. %So having analytic expressions for the lightcurves of basis maps makes the inverse problem faster to solve (Farr et al. 2018).    

Since planets are spherical, the basis maps of choice are spherical harmonics and the resulting lightcurves are called harmonic lightcurves. Alternatively, it is mathematically easy, and sometimes astronomically useful, to consider discrete albedo markings, e.g., pixels. The limiting case of infinitesimal pixels are $\delta$-functions. So in this paper we will also consider maps consisting of a uniformly dark planet with a single bright point, which we call $\delta$-maps.  

\cite{Russell_1906} derived analytic reflected lightcurves for asteroids with convex shapes observed at opposition (full phase).  The shape and albedo markings were parameterized via spherical harmonics. Cowan, Fuentes \& Haggard (2013; henceforth CFH13) derived reflected harmonic lightcurves\footnote{A ``harmonic lightcurve'' is the lightcurve induced by a spherical harmonic.} of synchronously-rotating spherical planets on edge-on orbits.  That case is relevant to near-term goals such as albedo mapping of short period planets, but is a far cry from a general solution to the reflected light forward problem.  

In this work, we derive general analytic solutions for the reflected light from a spherical planet.  We assume diffuse reflection (i.e., Lambertian scattering), but our solutions work regardless of orbital inclination, planetary obliquity, or spin rate. The mathematical crux of deriving harmonic lightcurves for arbitrary geometry is rotating from the planetary coordinate system of the map to the coordinate system centered on the lune-shaped convolution kernel (Figure~\ref{PlanetGeometry}). Fortunately, rotations of spherical harmonics are useful in other areas of Physics and are compactly described by Wigner $D$-matrices. 
The bottom line is that we can express a harmonic lightcurve for an arbitrary viewing geometry as a non-linear combination of harmonic lightcurves for the simpler viewing geometry presented in CFH13.

We have implemented the analytic results of this paper in an open-source Mathematica code, the Exoplanet Analytic Reflected Lightcurves package, EARL (\url{github.com/HalHaggard/EARL}). EARL can be used as a black box or as a white box and makes lightcurve numerical experiments accessible. 

In Section~\ref{OrbGeoms} we describe the orbital and viewing geometry of the general problem. In Section~\ref{delta-maps} we consider $\delta$-maps, which admit a mathematically trivial analytic lightcurve. Despite being a good stand-in for small pixels, $\delta$-maps are an overcomplete basis, and hence of limited use for solving inverse problems. In Section~\ref{Ylm-maps} we therefore tackle the more useful---and mathematically challenging---case of harmonic lightcurves.

\begin{figure*}%  figure placement: here, top, bottom, or page
   \centering 
   \includegraphics[width=.45\textwidth]{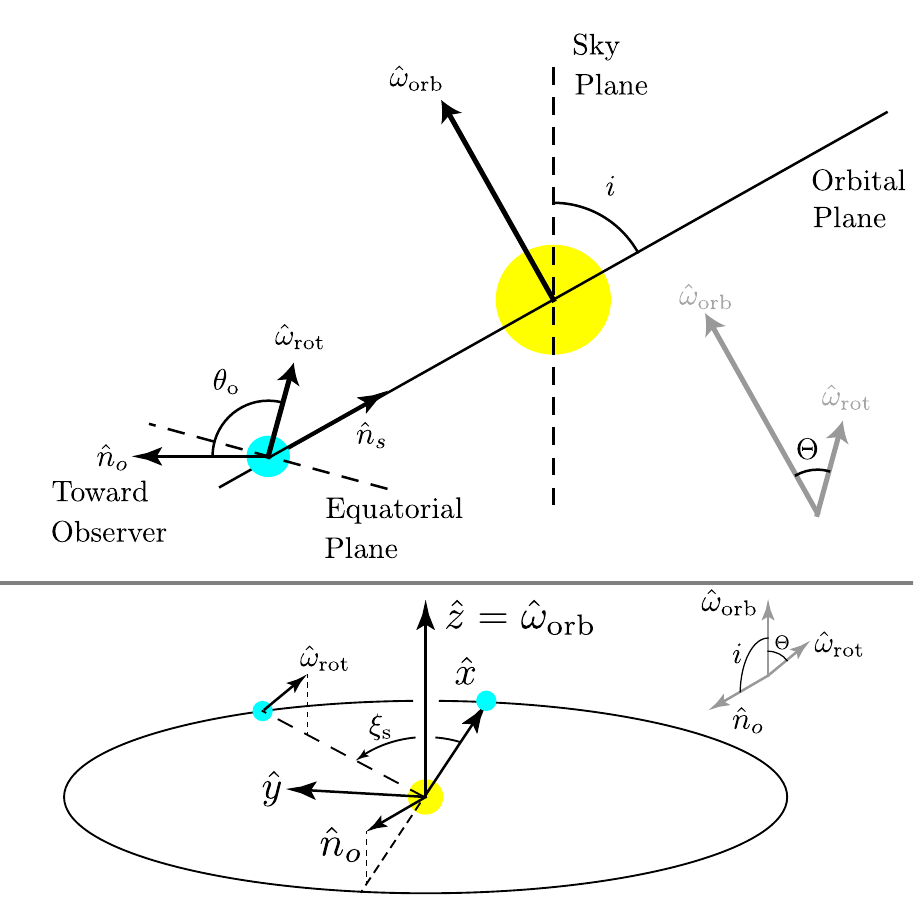} 
   \hspace{.5in}   
   \includegraphics[width=.45\textwidth]{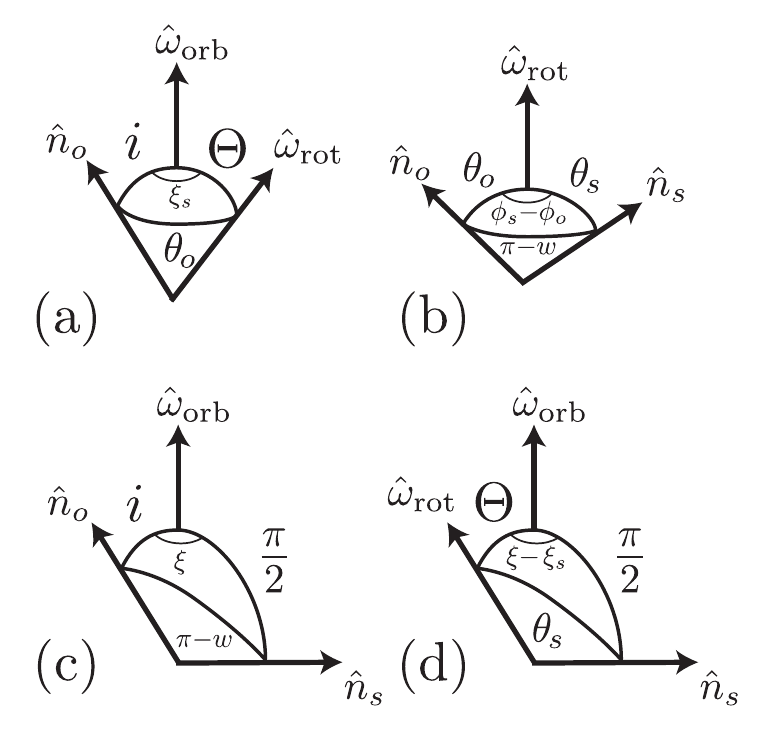} 
   \caption{This figure sets out conventions and angle relations for the rich geometry of an exoplanet on a circular orbit around a distant star. The upper left panel lays out the viewing geometry. The lower left panel gives the orbital geometry. The insets (a)-(d) of the right panel give the spherical angle relations amongst the various angles. }
   \label{OrbGeometry}
\end{figure*}

\section{Viewing and Orbital Geometries}
\label{OrbGeoms}

Here we summarize our conventions for the viewing and orbital geometries of a planet on a circular orbit about a distant star. The subobserver longitude $\phi_{o}(t)$  and the substellar angles $\theta_s(t)$ and $\phi_s(t)$ vary in time. \cite{Schwartz_2016} presented analytic expressions for these quantities as a function of seven system parameters, which we describe presently. We display the results and provide a new and more intuitive derivation in Appendix \ref{geometry}. Analytic expressions for the three time varying angles allow us to first find the harmonic lightcurves as functions of the subobserver and substellar points and only afterwards substitute their explicit time dependence, a particularly simple way to proceed.  

The first three system parameters are intrinsic to the system: rotational angular frequency, $\wrot \in (-\infty,\infty)$, orbital angular frequency, $\worb \in (0,\infty)$, and obliquity, $\Theta \in [0,\pi/2]$. Rotational frequency is measured in an inertial frame, where positive values are prograde with respect to the orbital motion and negative denotes retrograde rotation (for comparison, the rotational frequency of Earth is $\wrot^{\oplus}~\approx~2\pi/23.93\ \text{h}^{-1}$). Two more parameters are extrinsic and differ for each observer:  
orbital inclination, $i \in [0,\pi]$ where $i=0$ is face-on and orbiting counterclockwise as seen by the observer, $i=\pi/2$ is edge-on, and $i=\pi$ is face-on and orbiting clockwise,
and solstice phase, $\xis \in [0,2\pi)$, which is the orbital angle between superior conjunction and the maximum Northern excursion of the sub-stellar point (i.e., the orbital location of northern summer solstice). The remaining parameters are initial conditions: the starting orbital position, $\xi(0) \in [0,2\pi)$, and the initial sub-observer longitude, $\phobsz \in [0,2\pi)$. We adopt---with no loss of generality---the initial conditions $\phobs(0)=0$ and $\xi(0)=0$.  As a result, the sub-observer longitude and orbital location are simply given by $\phi_o = -\wrot t$ and $\xi = \worb t$, respectively. Figure \ref{OrbGeometry} shows these parameters in two different views on the left and four different sets of relations between them in the insets on the right. The analytic expression for the time-varying angles $\phi_{o}(t)$, $\theta_{s}(t)$, and $\phi_{s}(t)$ are given in Eqs. (\ref{eq:c_thobs}-\ref{eq:s_phst}).

\begin{figure*}%  figure placement: here, top, bottom, or page
   \centering
   \includegraphics[width=.40\textwidth]{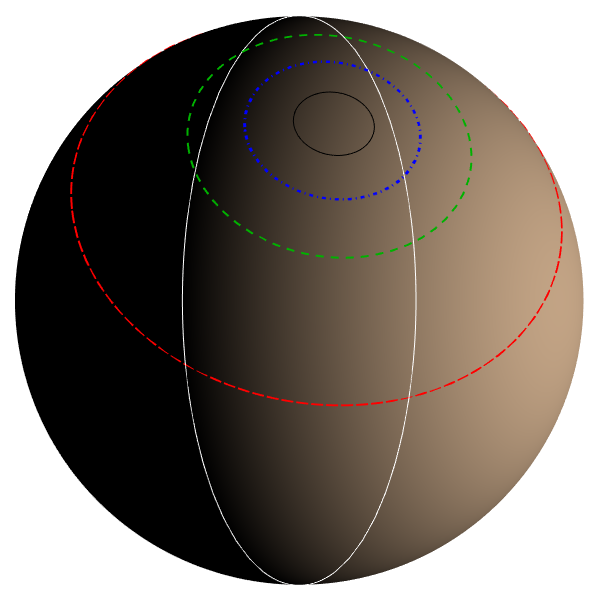} 
   \includegraphics[width=.55\textwidth]{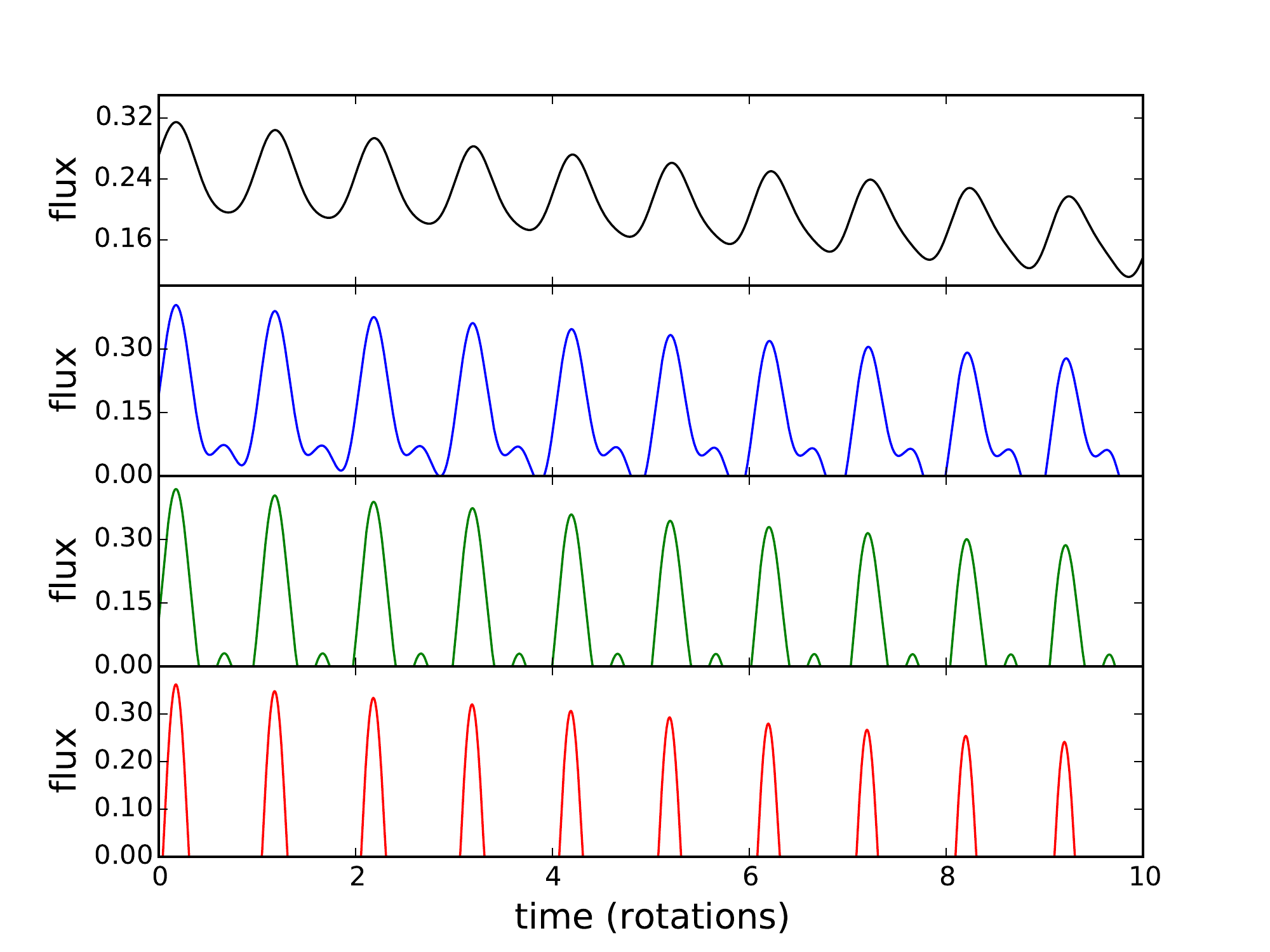} 
   \caption{\emph{Left:} Even at fixed orbital phase the $\delta$-maps exhibit a rich set of interactions with the illuminated and visible lune. The five qualitatively distinct interactions are depicted in this figure for a fixed relative position of planet and lune poles. A $\delta$-map at small colatitude remains within the lune throughout the rotation (solid). At middling colatitudes the spot exits and returns through one side of the lune (dot-dashed) or exits and returns through both sides of the lune (short dash). At still larger colatitudes the spot traverses the lune once (long dash). Finally, for colatitudes larger than that of the lune south pole the spot map is outside the lune for the whole rotation and gives no light curve, putting it in the nullspace. %For clarity we have depicted this case with a different choice of planet pole (dotted curve).
   \emph{Right:} Example lightcurves for 4 spots at the same planetary longitude but different latitudes, which pass through the kernel as shown in the left panel.} 
   \label{SpotMapTypes}
\end{figure*}

\section{The Simplest Basis: $\delta$-maps}\label{delta-maps}
Arguably the simplest orthonormal basis maps are delta functions: $M(\theta,\phi)= \delta(\theta'-\theta, \phi'-\phi)/\sin\theta'$, where $\theta'$ and $\phi'$ are the usual co-latitude and longitude on the planet, and $\theta$ and $\phi$ are the location of the bright point (the map is black everywhere else).

Such maps have the desirable property that they are easy to integrate so one can trivially compute the light curve for a $\delta$-map. Using the standard spherical measure $d \Omega' \equiv \sin \theta' d\theta' d\phi'$ this light curve is:  
\begin{equation} \label{delta_map}
\begin{aligned}
F_{\theta\phi}(t) = & \oint K(\theta', \phi', t) \frac{1}{\sin\theta'} \delta(\theta'-\theta, \phi'-\phi)d\Omega' \\
= & \int_0^\pi \int_0^{2\pi} K(\theta', \phi', t) \delta(\theta'-\theta, \phi'-\phi) d\theta' d\phi'  \\
= & K(\theta, \phi, t).
\end{aligned}
\end{equation}
In other words, the lightcurve for a $\delta$-function at location $(\theta,\phi)$ is simply the changing value of the kernel at that location. This is much faster to compute than numerically integrating the varying flux from a small pixel.

In the case of reflected light from a Lambertian reflector, the convolution kernel is the normalized product of visibility and illumination:
\begin{equation}
\label{kernel}
K(\theta, \phi, t) = \frac{1}{\pi} \max\big( V_{\rm nz}(\theta, \phi, t), \hspace{0.1cm} 0 \big) \cdot \max\big( I_{\rm nz}(\theta, \phi, t), \hspace{0.1cm} 0 \big),
\end{equation}
where the non-zero regions of the visibility and illumination are:
\begin{equation}
V_{\rm nz}(\theta, \phi, t) = \sin \theta \sin \theta_{o} \cos(\phi-\phi_{o}) + \cos \theta \cos \theta_{o},
\end{equation}
\begin{equation}
\label{illum}
I_{\rm nz}(\theta, \phi, t) = \sin \theta \sin \theta_{s} \cos(\phi-\phi_{s}) + \cos \theta \cos \theta_{s}.
\end{equation}

By substituting Equations (\ref{eq:c_thobs}-\ref{eq:s_phst}) into Equations (\ref{delta_map}-\ref{illum}), one obtains the explicit time-dependence for the kernel at a given location, or---equivalently---the lightcurve for a delta function at that location.  

In Figure~\ref{SpotMapTypes} we use $\delta$-maps to explore the different lightcurve morphologies. As described in detail below, the non-zero portion of the kernel forms a geometrical lune. The image on the left of Figure~\ref{SpotMapTypes} is centered on a lune, with the star is to the right, and the observer is to the left.  Depending on the location of the bright spot, there are four classes of behavior (excluding spots that don't rotate through the lune, which are in the nullspace).  

The polar spot in Figure~\ref{SpotMapTypes} (the top panel on the right) is continuously illuminated and visible: the flux from the spot never drops to zero during the plotted interval, but varies periodically. The second panel shows a spot somewhat farther from the pole.  At the start of the plotted interval, the spot remains within the lune throughout a planetary rotation but passes through a local maximum twice per rotation, hence the lightcurve exhibits a double-peaked morphology.  As the lune moves with respect to the planetary pole, this spot periodically leaves the lune. The third panel shows a spot at mid-latitudes that passes in and out of the lune twice per rotation. The fourth panel shows the lightcurve for a spot close to the equator: it passes through the lune once per rotation. 

At a given orbital phase, the shape of the kernel is fixed, as is its position with respect to the planetary axis of rotation, so the contribution of a region to rotational variability is solely determined by the latitude of that location.  But as the planet moves in its orbit, the shape and location of the kernel changes, and hence the lightcurve contribution from a given latitude can change character (continuously visible, double-peaked, single peaked, etc.). Lastly, note that due to the angle between the lune and the planetary coordinates, spots at the same longitude can contribute to the lightcurve at different rotational phases. The bottom line is that even the simplest albedo maps produce surprisingly varied lightcurves.

\section{Harmonic Lightcurves}\label{Ylm-maps}

In this section we give a full analytic solution to the problem of finding the lightcurve for a planet described by a spherical harmonic albedo map with an arbitrary viewing geometry.  As noted by CFH13, the map-to-lightcurve transformation is essentially a convolution, and the crux is that the kernel is piecewise-defined.  The non-zero part of the integral has limits related to the viewing geometry described in Section \ref{OrbGeoms}.  These limits are usually awkward: they do not correspond to constant latitude or longitude.  

However, there is always a coordinate system based on the terminator and the limb where the integral kernel---or equivalently the limits of integration---are as simple as possible. We call this system lune coordinates and describe how to transform from the planet-based to lune coordinates in subsection \ref{LuneGeom}. Subsection \ref{Result} uses the rotation properties of spherical harmonics to derive the light curve of an arbitrary spherical harmonic map and subsection~\ref{checking} compares these results to previous work. In the final subsection, \ref{Symm}, we find the full set of symmetries and derive some new recursion relations for harmonic lightcurves; using these symmetries allows substantial simplification in the computation of the full set of harmonic lightcurves.

\begin{figure*}%  figure placement: here, top, bottom, or page
   \centering
   \includegraphics[width=.43\textwidth]{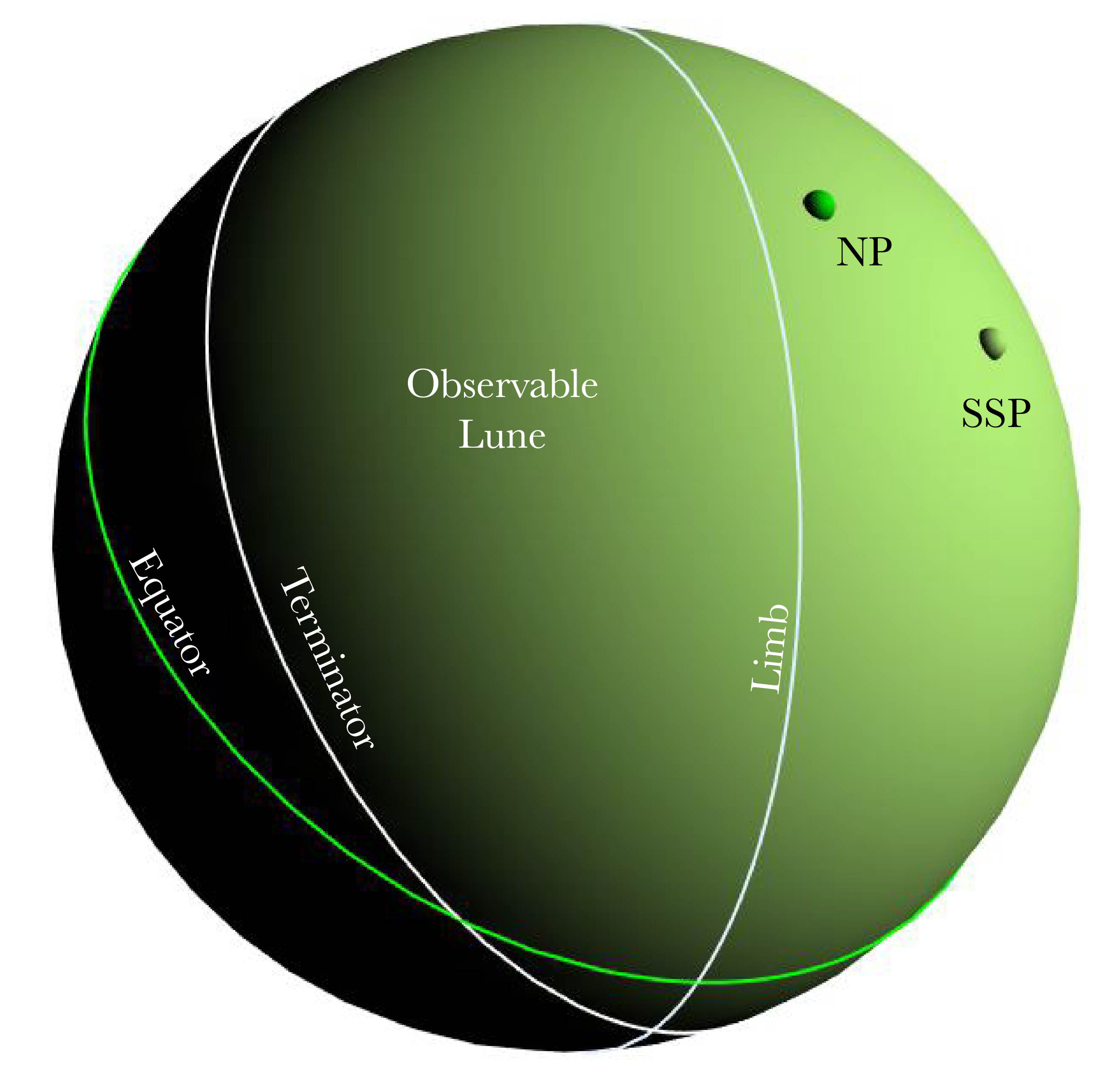} 
   \hspace{.04\textwidth}
   \includegraphics[width=.43\textwidth]{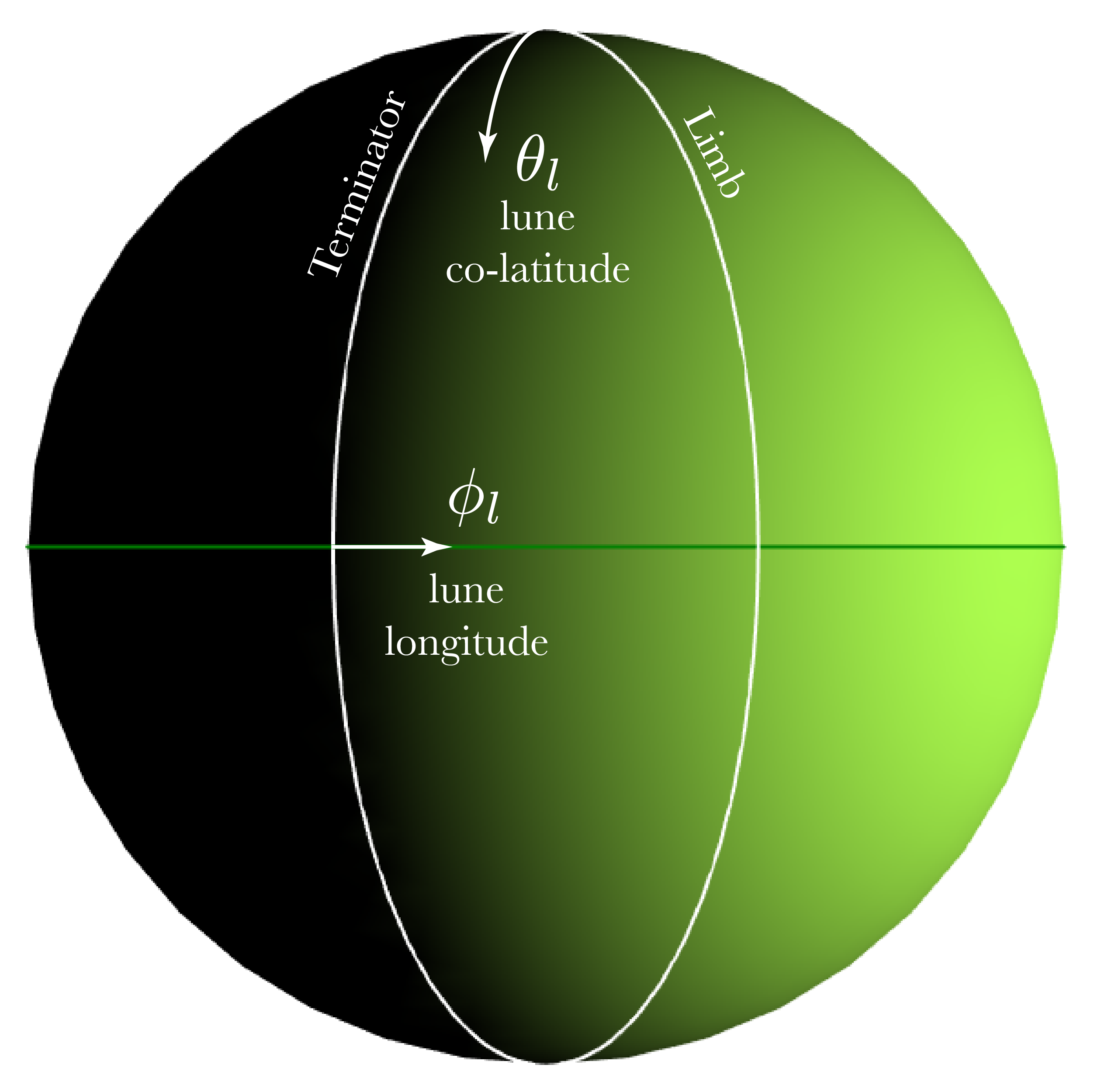} 
   \caption{\emph{Left:} No matter where the subobserver (SOP) and substellar (SSP) points fall on the planet, they define two great circles on the planet: the limb defines the boundary of the visible portion of the planet and the terminator defines the boundary of the illuminated portion. The region that is both visible and illuminated cuts out a wedge, a geometrical lune, on the planet and our lune coordinates are defined using this observable lune. The north pole (NP) generates a third great circle, the planetary equator. \emph{Right:} Rotation from the planet coordinates to the lune coordinates, illustrated in this figure, is the key step in being able to find an analytic solution to the forward problem for harmonic lightcurves. Lune coordinates use a co-latitude $\theta_l$, which descends from the north pole of the lune and a longitude $\phi_l$ with prime meridian set by the terminator. The angular width $w$ of the terminator-to-limb lune is a fundamental parameter in our analysis.}
   \label{PlanetGeometry}
\end{figure*}

\subsection{Lune Geometry}
\label{LuneGeom}

The general result presented here is based on the use of two different coordinates systems: one is the standard planetary coordinate system and is fixed by the rotation axis of the planet, the second is a coordinate system adapted to the instantaneous illumination and visibility of the planet (see Fig. \ref{PlanetGeometry}, left and right panels respectively). We will call the latter system lune coordinates, as it is in this coordinate system that the integral kernel is most simply expressed as a lune cut out by two great circles. The main purpose of this section is to establish a few conventions about the lune coordinates and to find the transformation from the planetary coordinates to the lune ones. 

The Cartesian coordinates of the subobserver and substellar points in the planet's coordinate system are defined by:
\begin{equation}
\begin{aligned}
\hat{n}_o &= (\sin \theta_o \cos \phi_o, \sin \theta_o \sin \phi_o, \cos \theta_o),\\
\hat{n}_s &= (\sin \theta_s \cos \phi_s, \sin \theta_s \sin \phi_s, \cos \theta_s).
\end{aligned}
\end{equation}
These two points provide the poles for the great circles that determine the limb and the terminator, respectively (Fig. \ref{PlanetGeometry}). 

The main issue of convention is that we need to fix an analog of the north pole for lune coordinates. The north pole of the planetary system is set by the axis of rotation and the right hand rule. We will fix the north pole of the lune coordinates to be
\begin{equation}
\hat{z}_{l} = \frac{\hat{n}_o \times \hat{n}_s}{\sin(\pi- w)} = \frac{\hat{n}_o \times \hat{n}_s}{\sin w},
\end{equation}
and call it the lune pole. Here $w$ is the angular width of the kernel; an explicit formula for it, in terms of the subobserver and substeller points, is given below. If we further fix the prime meridian of the lune coordinates to coincide with the terminator, then we can complete this north pole to a full  Cartesian coordinate frame with 
$\hat{y}_l = \hat{n}_s$ and $\hat{x}_l = \hat{y}_l \times \hat{z}_l$. An advantage of these choices is that the lune coordinates are only ill-defined when the substellar and subobserver points coincide or are antipodal, but both of these situations correspond to syzygy (either the planet transiting its star or being occulted by it) and are observationally uninteresting for the reflected light curve.

Having specified the two coordinate frames it is now a simple matter to find the Euler angles that specify the rotation from one system to the other (the numerical superscripts are indicating which components of the unit vectors are used): 
\begin{equation} \label{Eulers}
\begin{aligned}
\tan \alpha  &=  \frac{  {\hat{z}}_l^{(2)}}{ {\hat{z}}_l^{(1)}} =  \frac{ \cos\theta_o \sin\theta_s \cos\phi_s  - \cos\theta_s  \sin\theta_o \cos\phi_o }{ \cos\theta_s  \sin\theta_o \sin \phi_o - 
 \cos\theta_o \sin \theta_s \sin\phi_s},\\
\cos \beta  &=  {\hat{z}}_l^{(3)} = \frac{\sin \theta_o \sin \theta_s \sin(\phi_s - \phi_o)}{\sin w}, \\
\tan \gamma  &= \frac{\phantom{-}\hat{y}_l^{(3)}}{-\hat{x}_l^{(3)}} =  -\frac{\cos \theta_s \sin w }{\cos\theta_o + \cos \theta_s \cos w },
\end{aligned}
\end{equation}
and their ranges are $\alpha \in [0, 2 \pi]$, $\beta \in [0, \pi]$, and $\gamma \in [0, 2\pi]$, respectively.  In numerical treatments the $\text{atan2}\,(y,x)$ function should be used to extract $\alpha$ and $\gamma$. Note carefully that we use proper Euler angles and a $z$-$y'$-$z''$ intrinsic rotation convention. These Euler angles are crucial to the analytic solution given below where they provide the arguments of the Wigner $D$-matrices connecting the general spherical harmonics to the ones in the particularly simple lune geometry.  

The width of the kernel can also be written in terms of the subobserver and substellar points
\begin{equation}
\label{width}
\cos (\pi-w)  = \cos \theta_o \cos \theta_s + \sin \theta_o \sin \theta_s \cos (\phi_s-\phi_o),
\end{equation}
which follows from the spherical law of cosines applied to the spherical triangle shown in inset (b) of Fig. \ref{OrbGeometry}.  The width takes values in the range $w \in [0, \pi]$ and so here the cosine is sufficient.

\subsection{The Result}
\label{Result}
A natural set of basis maps for planets are spherical harmonics, $Y_l^m$ (see Appendix~\ref{Recurrence} for the exact normalization and phase conventions). Each spherical harmonic has a lightcurve signature called a harmonic lightcurve, 
\begin{equation}
\label{def}
F_l^m = \oint K(\theta', \phi', \mathbb{G}) Y_l^m(\theta', \phi') d \Omega',
\end{equation}
here the integral is over planet coordinates. A harmonic light curve can also be thought of as the time-varying components of the integral kernel in the spherical harmonic basis.
Despite the fact that this integral is taken over the whole sphere, the result is complicated by the intricate piecewise definition of the integral kernel, Eqs. \eqref{kernel}-\eqref{illum}, in the planet's coordinate system due to the viewing geometry $\mathbb{G}$.  

We take advantage of the transformation properties of spherical harmonics to simplify this integral. Under a rotation $\mathcal{R}$---characterized by the Euler angles $\alpha$, $\beta$, and $\gamma$---from the planet coordinates to the lune coordinates, we have
\begin{equation}
Y_l^m(\theta', \phi') = \sum_{m'=-l}^{l} \left[D^{(l)}_{m m'}(\mathcal{R}) \right]^{*} Y_l^{m'}(\theta, \phi).
\end{equation}
Here $D^{(l)}_{m m'}(\mathcal{R}) = D^{(l)}_{m m'}(\alpha,\beta, \gamma)$ is Wigner's $D$-matrix (Wigner 1927). Substituting this expression into Eq. \eqref{def} and exchanging the order of integration and summation gives 
\begin{equation}
F_l^m = \sum_{m'=-l}^{l} \left[D^{(l)}_{m m'}(\mathcal{R}) \right]^{*} \oint K(\theta, \phi, \mathbb{G}_0) Y_l^{m'}(\theta, \phi) d \Omega,
\end{equation}
where $\mathbb{G}_0$ indicates the lune geometry and we have used the fact that the angular measure $d\Omega$ is rotationally invariant. 

In the lune geometry, where the prime meridian is aligned with the terminator, the kernel vanishes everywhere except for lune longitudes $\phi \in (0,w)$. Within that range, the kernel is non-zero and takes the simplified form
\begin{equation}
K_{nz}(\theta, \phi, \mathbb{G}_0) = \frac{1}{\pi} \sin^2\theta \sin \phi \sin (w - \phi).
\end{equation}
The product form of the kernel allows us to separate the integral into two pieces depending on $\theta$ and $\phi$ separately. Because the associated Legendre polynomial depends directly on $\cos \theta$ it is convenient to change the integral over the co-latitude to $x= \cos \theta$. Then 
\begin{equation}
\begin{aligned}
\oint K(\theta, \phi, \mathbb{G}_0) Y_l^{m'}(\theta, \phi) d \Omega = \mathcal{P}_{l}^{m'} \Phi_{m'}(w)
\end{aligned}
\end{equation}
where 
\begin{equation}
\label{Pconst}
\mathcal{P}_l^{m'} \equiv \frac{(-1)^m}{\pi} \sqrt{\frac{(2 l+1)}{4\pi}\frac{(l-m')!}{(l+m')!}} \int_{-1}^{1}(1-x^2) P_{l}^{m'}(x) d x ,
\end{equation}
and 
\begin{equation}
\label{PhiDef}
\Phi_{m'}(w) \equiv \int_0^{w} \sin \phi \sin(w-\phi) e^{i m' \phi} d\phi.
\end{equation}
The integral of Eq. \eqref{Pconst} gives an $l$- and $m'$-dependent number $\mathcal{P}_l^{m'}$ that can be found explicitly using recurrence properties of the associated Legendre polynomials, see Appendix \ref{Recurrence}. Note that we include the Condon-Shortley phase, $(-1)^m$, in our convention and make it explicit rather than including it in the definition of the Legendre polynomials. The complete solution for the harmonic light curves is then
\begin{equation}
\label{Light}
F_l^m = \sum_{m'=-l}^{l} \mathcal{P}_l^{m'} \left[D^{(l)}_{m m'}(\mathcal{R}) \right]^{*}  \Phi_{m'}(w) ,
\end{equation}
where the integral $\Phi_{m'}$, Eq. \eqref{PhiDef}, can be solved explicitly, yielding 
\begin{equation}
\label{Phi}
\Phi_{m'}(w) = \left\{ \begin{array}{cr} \frac{1}{2}(\sin w - w \cos w) & m'=0\\[6pt]
	 \frac{1}{4} e^{\pm i w} (w-\cos w \sin w) & m' = \pm 2 \\[6pt]
	 \frac{2i \cos w (1-e^{im' w})-m' \sin w (1+e^{im' w})}{m'(m'^2-4)} & \textrm{else.} \end{array} \right.
\end{equation}
Note that the summation in Eq. \eqref{Light} really only extends over half of the $m'$ values because the coefficients $\mathcal{P}_{l}^{m'}$ vanish when $l+m'$ is odd. Equation \eqref{Light}, together with \eqref{Eulers}, \eqref{Phi}, and \eqref{Pconsts}, gives a complete analytic formula for an arbitrary harmonic lightcurve. 

In the present work we include the Condon-Shortley phase, $(-1)^m$, in the definition of the spherical harmonics and work with complex spherical harmonics. This is consistent with a number of modern works in astrophysics and largely agrees with the default options in software packages, such as Python and Mathematica. This is however a different convention than in our previous works \citep[CFH13;][]{Schwartz_2016}. Appendix \ref{ComplexConventions} details how to recover the real conventions without the Condon-Shortley phase if needed. 

\subsection{Checking the Result}\label{checking}
Once the right setup has been identified, the analytic derivation above involves few steps. Nonetheless, there are a number of conventions that go into it and it is important to check the result of Eq. \eqref{Light} against previous work. 

We begin with the simple $l=0$ case. In this case the Wigner $D$-matrix equals one, $D^{(0)}_{00} = 1$, independent of its arguments, and the coefficient is $\mathcal{P}_0^{0} = 2/3\pi^{3/2}$,  so 
\begin{equation}
\label{Lambert}
F_0^{0} = \frac{1}{3\pi^{3/2}}(\sin w- w \cos w).
\end{equation}
This is the canonical Lambert phase curve \citep{Russell_1916}, but expressed in terms of the lune width $w$. 
For an edge-on and tidally locked orbit this lightcurve can be expressed in terms of the star--planet--observer phase angle, $\varphi=\pi-w$. 
Other normalization conventions are also common and we describe the conversion to that of CFH13 in Appendix \ref{Recurrence}.

\subsubsection{Edge-on and tidally locked orbit}

The edge-on and tidally locked scenario nicely highlights a subtlety of the analytic formula \eqref{Light}. Without loss of generality we can take the prime meridian of the planet coordinates to agree with the substellar longitude in this case, $\phi_s = 0$. Simplifying the formula for $\beta$, Eq. \eqref{Eulers}, using $\theta_s=\theta_o=\pi/2$, $\phi_s=0$, and Eq. \eqref{width} we find
\begin{equation}
\cos\beta = - \sin\phi_o/\sqrt{\sin^2\phi_o} = - \text{sgn}(\phi_o),
\end{equation}
where $\text{sgn}\, (x)$ returns $-1$ if $x$ is negative and $+1$ if $x$ is positive. This means that it is essential to take $\phi_o \in [-\pi,\pi]$ and that $\beta$ is zero while the planet orbits from superior conjunction to inferior conjunction, but $\beta=\pi$ while the planet orbits from inferior to superior conjunction. This discontinuity is due to the fact that the lune pole discontinuously flips from the planet's North pole to its South pole at inferior conjunction as the limb and terminator pass through each other (and similarly at superior conjunction). The analytic formulas capture this transition perfectly, but care is required in evaluating the edge-on, tidally locked case. From superior conjunction to inferior one can use $(\alpha, \beta, \gamma)=(0, 0,3\pi/2)$ and from inferior to superior conjunction one can use $(\alpha,\beta,\gamma)=(0,\pi, \pi/2)$.  Here we have arbitrarily resolved the gimbal lock ambiguity by putting all of the $z$-axis rotation into $\gamma$, but splitting it between $\gamma$ and $\alpha$ or putting it all into $\alpha$ are also valid choices. 

With this subtlety noted, Eq. \eqref{Light} gives lightcurves for the edge-on and tidally locked reflected light case in perfect agreement with CFH13.\footnote{There was a typo in CFH13: there $F_2^{-2}$ should have been $2 F_2^{-2}$. We have confirmed this numerically as well.} Notice that in this case the rotation from the planet to the lune coordinates is constant except for at inferior and superior conjunction, when the  lune pole switches discontinuously: the prime meridian of the lune coordinates is the day-night-terminator, which is stationary on a synchronously-rotating planet. This shows that all of the time-dependence of the lightcurve except for a sign is due to the time-varying lune width $w$, a result present but not highlighted in CFH13.  The discontinuous changes of the lune pole also explain why so many of the edge-on and tidally locked lightcurves of CFH13 can be expressed in terms of the star-planet-observer phase angle $\varphi=\pi-w$ alone. 

\subsubsection{Tidally locked and inclined orbit}

In the tidally locked and inclined case we can again take $\theta_s = \pi/2$ and $\phi_s=0$ and two of the Euler angles take on special values. Evaluating Eq. \eqref{Eulers} carefully using $\text{atan2}\,(y,x)$ yields $\alpha = \pi/2$, $\gamma = \pi$, and $\cos \beta = -\sin \theta_o \sin \phi_o/\sqrt{1-\sin^2 \theta_o \cos^2 \phi_o}$. Using these Euler angles in Eq. \eqref{Light} we are able to reproduce the numerically computed inclined lightcurves of CFH13 analytically. We have tabulated several of these new analytic lightcurves along with an example of an arbitrary viewing geometry lightcurve in Appendix \ref{Explicit}.

\subsubsection{Thermal light}

We conclude this section with a comment on the case of thermal light. In CFH13 the case of arbitrary thermal harmonic lightcurves was reduced to a set of integrals that was particularly tractable and several of the low order harmonic lightcurves were tabulated. However, no general analytic solution to these integrals was found. The methods used in this paper can easily be adapted to the case of thermal light as well. Note that the kernel for planetary emission is different than for reflected light so one cannot simply use Eq. \eqref{Light}. Nonetheless, it is a straightforward task to repeat our argument for the case of a kernel that only includes visibility.  

\subsection{Symmetries and Lightcurve Recursion}
\label{Symm}

The (complex) harmonic lightcurves have the simple symmetry property \begin{equation}
\label{symm}
F_{l}^{-m} = (-1)^m [F_{l}^{m}]^*.
\end{equation}
 This symmetry follows immediately from the definition, Eq. \eqref{def}, and the symmetry, $Y_{l}^{-m} = (-1)^m [Y_{l}^m]^*$, of the spherical harmonics:
\begin{equation}
\label{derivsymm}
\begin{aligned}
F_{l}^{-m} &= (-1)^m \oint K(\theta',\phi',\mathbb{G}) [Y_{l}^{m}(\theta',\phi')]^* d\Omega'\\
&=(-1)^m \left[\oint K(\theta',\phi',\mathbb{G}) Y_{l}^{m}(\theta',\phi') d\Omega' \right]^*\\
&=(-1)^m [F_{l}^m]^*.
\end{aligned}
\end{equation}
Hence it is only necessary to compute half of the lightcurves, say those with $m\ge 0$, and the others follow immediately.

The Wigner $D$-matrices also have extensive symmetries and satisfy recursion relations \citep[for a detailed compendium see][]{Varshalovich_1988}. Intriguingly, almost all of these properties depend on both the $m$ and $m'$ indices. Because our main result is a sum over $m'$, introducing new $m'$-dependent terms makes it difficult to do the sum explicitly and only a few of these properties can be extended to the harmonic lightcurves. For example, the symmetry of Eq. \eqref{symm} can also be more laboriously proved using the symmetries of the $D$-matrices. In the remainder of this section we derive a recursion relation for the harmonic lightcurves using properties of the $D$-matrices. Whether further relations can be derived along these lines is an interesting subject for future study.

A basic recursion relation for the $D$-matrices is
\begin{equation}
\frac{2m'-2m \cos \beta}{\sin \beta} D^{(l)}_{m m'} = L_{-}D^{(l)}_{m-1 m'} e^{-i \alpha}+ L_{+} D^{(l)}_{m+1 m'} e^{i \alpha},
\end{equation}
where the coefficients are $L_{\pm} \equiv
 \sqrt{(l\mp m)(l \pm m+1)}.$
Taking the complex conjugate of both sides of this recursion, multiplying by $\mathcal{P}_{l}^{m'} \Phi_{m'}(w)$, summing over $m'$, and using the differential relation 
\begin{equation}
\label{Deriv}
\frac{\partial}{\partial \gamma} D^{(l)}_{mm'} = -i m' D^{(l)}_{m m'},
\end{equation}
 we obtain the following differential recursion relation for the harmonic lightcurves:
\begin{equation}
\label{recurs}
\left(\frac{-2i}{\sin \beta} \frac{\partial}{\partial \gamma} -2m \cot \beta\right) F_{l}^{m} = L_{-}F_{l}^{m-1} e^{i \alpha}+ L_{+} F_{l}^{m+1} e^{-i \alpha} .
\end{equation}
Because the result no longer involves a summation, this provides an efficient strategy for obtaining all the harmonic lightcurves with $m<l$ from $F_l^l$ provided one is working analytically and can evaluate the derivative. Of course, the process is easily automated in a framework like Mathematica. We illustrate this recursion along with several explicit examples of harmonic lightcurves in Appendix \ref{Explicit}.

This derivation requires the technical assumption that the $\gamma$ derivative of Eq. \eqref{Deriv} can be pulled outside of the sum. Since the only thing that depends on $\gamma$ is the $D$-matrix this seems to be valid. However, since the Euler angles and the width $w$ all depend on the subobserver and substellar points, it could be that not all four quantities are independent and hence that the derivative cannot be taken at fixed $w$. In Appendix \ref{Jacobian}, we address this issue by showing that the Jacobian from $\{\alpha, \beta, \gamma, w\}$ to $\{\theta_s,\phi_s,\theta_o, \phi_o\}$ is well behaved.

\section*{Acknowledgments}
We thank the International Space Science Institute in Bern, Switzerland, for hosting the Exo-Cartography workshop series. We used the ReflectDirect package created by J.C.~Schwartz. HMH thanks the IGC at Pennsylvania State University for warm hospitality while completing this work, Bard College for extended support to visit the ISSI with students, and the Perimeter Institute for Theoretical Physics for generous sabbatical support. This work is  supported  by  Perimeter  Institute  for  Theoretical  Physics.   Research  at  Perimeter  Institute is supported by the Government of Canada through Industry Canada and by the Province of Ontario through the Ministry of Research and Innovation.

\bibliography{CatsEye}

\onecolumn
\appendix

\section[]{Spherical Harmonics,  Legendre Polynomials, and Their Integrals}
\label{Recurrence}

Despite the standard nature of the spherical harmonics, we find it useful to explicitly display some low order spherical harmonics; this allows rapid comparison of phase and normalization conventions, see Table \ref{Ytab}. The spherical harmonics are
\begin{equation}
Y_{l}^{m}(\theta, \phi) = (-1)^m \sqrt{\frac{(2 l+1)}{4\pi} \frac{(l-m)!}{(l+m)!}} P_{lm}(\cos \theta) e^{im \phi},
\end{equation}
where $P_{lm}$ denotes the associated Legendre polynomial and we have made the Condon-Shortley phase, $(-1)^m$, explicit. Note carefully that this phase is often incorporated in the definition of the Legendre polynomials, e.g. in Mathematica, and should not be included twice. 

\begin{table}
   \centering
   %\topcaption{Table captions are better up top} % requires the topcapt package
   \begin{tabular}{@{} lccccc @{}} % Column formatting, @{} suppresses leading/trailing space
      \toprule
      $l$&\multicolumn{5}{c}{\phantom{spaces}$m$} \\
      \midrule
      $0$      & &  & $Y_0^0 = \frac{1}{2\sqrt{\pi}}$ &  & \\
      $1$      & &  $Y_1^{-1} = \frac{1}{2} \sqrt{\frac{3}{2\pi}} \sin \theta e^{- i \phi}$ & $Y_1^0 = \frac{1}{2} \sqrt{\frac{3}{\pi}} \cos \theta $ &  $Y_1^1 =  -\frac{1}{2} \sqrt{\frac{3}{2\pi}} \sin \theta e^{ i \phi}$ & \\
      $2$      & &  $Y_2^{-1} = \frac{1}{2} \sqrt{\frac{15}{2\pi}} \sin \theta \cos \theta e^{- i \phi}$ & $Y_2^0 =\frac{1}{4} \sqrt{\frac{5}{\pi}} ( 3\cos^2 \theta -1) $ &  $Y_2^1 = -\frac{1}{2} \sqrt{\frac{15}{2\pi}} \sin \theta \cos \theta e^{ i \phi}$ & \\
      		&	&$Y_2^{-2} =\frac{1}{4} \sqrt{\frac{15}{2\pi}} \sin^2 \theta e^{- i 2\phi} $  \phantom{TextSpac}& & \phantom{TextSpac}$Y_2^{2} =\frac{1}{4} \sqrt{\frac{15}{2\pi}} \sin^2 \theta e^{ i 2\phi} $ & \\
      \bottomrule
   \end{tabular}
   \caption{The low order spherical harmonics for $l \in \{0,1,2\}$. We adopt the conventions most often used in quantum mechanics. That is, we use complex spherical harmonics that are orthonormal with respect to the standard spherical measure, include the Condon-Shortley phase, and adopt colatitude as our $\theta$ coordinate.}
   \label{Ytab}
\end{table}

\subsection{Real and Complex Conventions for Spherical Harmonics and Harmonic Lightcurves}
\label{ComplexConventions}

In CFH13 we used conventions akin to those in geodesy with a real basis of spherical harmonics and did not include the Condon-Shortley phase. Thus in CFH13 the real spherical harmonics $\tilde{Y}_l^m$ we used were:
\begin{equation}
\tilde{Y}_{l}^m = \begin{cases}
\begin{aligned}
&N_l^m P_{l m}(\cos \theta) \cos(m \phi) \quad &\text{if $m\ge 0$}\\
&N_l^{|m|} P_{l|m|}(\cos \theta) \sin(|m| \phi) \quad &\text{if $m < 0$}
\end{aligned}
\end{cases} \qquad \text{with} \qquad 
{N}_{l}^m = \begin{cases}
\begin{aligned}
& 1 \quad &\text{if $l = 0$}\\
& \sqrt{\frac{2(2l+1)(l-m)!}{(l+m)!}} \quad &\text{if $l > 0$}
\end{aligned}
\end{cases}.
\end{equation}
That paper's normalization convention was a bit unusual as it included a factor of $\sqrt{2}$ even for $m=0$. Expressions for the real spherical harmonics $\tilde{Y}_{l}^m$ in terms of of the complex ones used in this paper $Y_{l}^m$ can be used to recover the CFH13 results from those in this paper. As shown in Eqs. \eqref{symm} and \eqref{derivsymm}, the complex symmetries of the harmonic lightcurves are the same as those of the spherical harmonics. For the reader's ease we report the real harmonic lightcurves $\tilde{F}_l^m$ in terms of the complex ones, $F_l^m$, presented in this paper. All real lightcurves are given in the conventions of CFH13, detailed just above: 
\begin{equation}
\tilde{F}_{l}^m = \begin{cases}
\begin{aligned}
(-1)^m  \sqrt{2 \pi} \left[ {F}_l^m+(-1)^m {F}_l^{-m}\right] \quad &\text{if $m\ge 0$ and $l > 0$},\\
\sqrt{4 \pi} {F}_0^{0} \hspace{.8in}  &\text{if $m=0$ and $l=0$},\\
-i \sqrt{2 \pi} \left[ (-1)^m {F}_l^{-m}-{F}_l^m \right] \quad &\text{if $m<0$}.
\end{aligned}
\end{cases}
\end{equation}

\subsection{Legendre Polynomial Recurrence Relation}
Recurrence relations allow us to relate associated Legendre polynomials with different degrees and orders. Building off of the relations in \cite{DiDonato_1982} we obtain
\begin{equation}
\label{DiDonato}
\sqrt{1-x^2} P_{l}^m =  \frac{1}{2l+1} \left[ P_{l+1}^{m+1}- P_{l-1}^{m+1}  \right].
\end{equation}
Note that \cite{DiDonato_1982} also does not include the Condon-Shortley phase in the definition of the associated Legendre polynomials. The above equation, hence, differs by a minus sign from Mathematica and other sources that use this phase, however, the recursion relation below is unaffected by this choice because it relates polynomials for which the $m$-indices differ by two. Also, the appearance of the pair $(l-1, m+1 )$ should inspire caution, however it does not present any difficulties as $P_{l-1}^{m+1}$ simply vanishes when $m+1> l-1$ and the recurrence continues to hold. Iterating the recurrence \eqref{DiDonato} we obtiain,
\begin{equation}
\label{iter}
(1-x^2) P_{l}^{m} = \frac{1}{4 l^2 -1} \left( \frac{2l-1}{2 l+3} P_{l+2}^{m+2} -\frac{4l+2}{2l+3} P_l^{m+2} + P_{l-2}^{m+2} \right).
\end{equation}

\subsection{Definite Integrals of the associated Legendre polynomials on $x \in [-1,1]$ and the coefficients $\mathcal{P}_l^m$}
For the special case where the limits of integration are $x \in [-1,1]$, compact solutions (i.e., not involving sums) for the definite integral of the associated Legendre polynomials have been worked out by \cite{Jepsen_1955}: 
\begin{equation}\label{Jepsen_Eqn}
R_l^m \equiv \int_{-1}^{1} P_{lm}(x) dx  =  \left\{ \begin{array}{cr} 	 2 & \textrm{if $l=m=0$}\\[6pt]
R_l^m(\textrm{even}) & \textrm{if $l$ and $m$ are even}\\[6pt]
	 R_l^m(\textrm{odd}) & \textrm{if $l$ and $m$ are odd}\\[6pt]
	 0 & \textrm{if $l+m$ is odd,} \end{array} \right.
\end{equation}
where the $R$ functions are
\begin{equation}
\label{Rs}
\begin{aligned}
R_l^m(\textrm{even}) &\equiv \frac{2|m|[(l/2)!]^2(l+m)!}{l[(l-m)/2]! [(l+m)/2]! (l+1)!}, \\
R_l^m(\textrm{odd}) &\equiv\frac{-\pi m(l+m)! (l+1)!}{l 2^{2l+1} \{[(l+1)/2]!\}^2 [(l-m)/2]! [(l+m)/2]!}.
\end{aligned}
\end{equation}
Physically, the northern and southern hemispheres have perfectly canceling light curves in the fourth case. 

\cite{Jepsen_1955} do not explicitly treat the case of $m<0$, so we check directly
\begin{equation}
R_l^{-m} \equiv \int_{-1}^{1} P_{l -m}(x) dx = \int_{-1}^{1} (-1)^{m}\frac{(l-m)!}{(l+m)!} P_{lm}(x) dx =  (-1)^{m} \frac{(l-m)!}{(l+m)!} R_l^{m}.
\end{equation}
This shows that we can put $-m$ into the Eqs. \eqref{Rs} and get the correct answer. The sign $(-1)^m$ in this formula is independent of the Condon-Shortley phase and comes only from the Rodrigues' formula for the associated Legendre polynomials. 

The coefficients $\mathcal{P}_l^m$ appearing in our main result, Eq. \eqref{Light}, are defined by
\begin{equation}
\mathcal{P}_l^m \equiv  \frac{(-1)^m}{\pi} \sqrt{\frac{(2 l+1)}{4\pi}\frac{(l-m')!}{(l+m')!}} \int_{-1}^{1}(1-x^2) P_{l}^{m'}(x) d x.
\end{equation}
 Using the recurrence \eqref{iter} and the results for the total integrals of the Legendre polynomials, we find that these coefficients can be expressed as
\begin{equation}
\label{Pconsts}
\mathcal{P}_{l}^m = \left\{ \begin{array}{cr} \frac{1}{\pi(4 l^2 -1)} \sqrt{\frac{(2 l+1)(l-m)!}{(l+m)!}} \left( \frac{2l-1}{2 l+3} R_{l+2}^{m+2} -\frac{4l+2}{(2l+3)} R_l^{m+2} + R_{l-2}^{m+2} \right) & m\le l -4,\\[6pt]
\frac{1}{\pi(4 l^2 -1)} \sqrt{\frac{(2 l+1)(l-m)!}{(l+m)!}} \left( \frac{2l-1}{2 l+3} R_{l+2}^{m+2} -\frac{4l+2}{(2l+3)} R_l^{m+2}  \right) & l -4 < m\le l -2,\\[6pt]
\frac{1}{\pi(4 l^2 -1)} \sqrt{\frac{(2 l+1)(l-m)!}{(l+m)!}} \left( \frac{2l-1}{2 l+3} R_{l+2}^{m+2}  \right) & l-2 <m\le l. \end{array} \right.
\end{equation}

\section{Phase convention for the $D$-matrices}
Our phase convention for the $D$-matrices is such that
\begin{equation}
D^{(l)}_{m m'}(\alpha, \beta, \gamma) = e^{-i \alpha m} d^{(l)}_{m m'}(\beta) e^{-i \gamma m'},
\end{equation}
where $d^{(l)}_{mm'}$ is the Wigner small $d$-matrix and is given by the single sum formula
\begin{equation}
d^{(l)}_{mm'}(\beta)  = (-1)^{l+m}[(l+m)!(l-m)!(l+m')!(l-m')!]^{1/2} \sum_{k} (-1)^k \frac{\left( \cos \frac{\beta}{2} \right)^{2k-m-m'} \left( \sin \frac{\beta}{2} \right)^{2l+m+m'-2k}}{k!(l+m-k)!(l+m'-k)!(k-m-m')!},
\end{equation}
as well as many other expressions. The summation index $k$ takes all values such that the arguments of the factorials are non-negative. Our conventions agree with  \cite{Varshalovich_1988}, but differ from, for example, Mathematica, which uses $\{\alpha,\beta,\gamma\} \rightarrow \{-\alpha, -\beta, -\gamma\}$.  Again, we use proper Euler angles and a $z$-$y'$-$z''$ intrinsic rotation convention.  

\section{Derivation of the Time-Varying Geometry}\label{geometry}
As discussed in Section \ref{OrbGeoms}, it is simplest to first find the harmonic lightcurves as a function of the substellar and subobserver points and only afterwards substitute the time dependence of these points to find the fully time-dependent lightcurve. Here we present simplified derivations for the subobserver colatitude in terms of the parameters of the viewing and orbital geometry and the explicit time dependence of $\phi_{o}(t)$  and the substellar angles $\theta_s(t)$ and $\phi_s(t)$. The resulting expressions are:\footnote{There was an obvious typo in the expression for $\cos\phi_o$ in Schwartz et al. (2016).}
\begin{equation}
%\begin{aligned}
	\label{eq:c_thobs}
	\cos\thobs = \cos i \cos \Theta + \sin i \sin\Theta \cos\xi_s,
    \qquad
    \sin\thobs = \sqrt{1-\cos^2\thobs},
%    \end{aligned}
\end{equation}
\begin{equation}
	\label{eq:c_phobs}
	\cos\phobs = \cos(\wrot t),
\qquad
	\sin\phobs = -\sin(\wrot t),
\end{equation}
\begin{equation}
%\begin{aligned}
	\label{eq:c_thst}
	\cos\thst = \sin\Theta\cos(\worb t - \xis),
    \qquad
	\sin\thst = \sqrt{1 - \sin^{2}\Theta\cos^{2}(\worb t - \xis )},
    %\end{aligned}
\end{equation}
\begin{equation}
	\label{eq:c_phst}
	\cos\phst = \frac{\cos(\wrot t)\Big[\sin i\cos(\worb t) - \cos\thobs\sin\Theta\cos(\worb t - \xis)\Big] + \sin(\wrot t)\Big[\sin i\sin(\worb t)\cos\Theta - \cos i\sin\Theta\sin(\worb t - \xis )\Big]}{\sqrt{1 - (\cos i \cos \Theta + \sin i \sin\Theta \cos\xi_s)^2}\sqrt{1 - \sin^{2}\Theta\cos^{2}(\worb t - \xis)}},
\end{equation}
\begin{equation}
	\label{eq:s_phst}
	\sin\phst = \frac{-\sin(\wrot t)\Big[\sin i\cos(\worb t) - \cos\thobs\sin\Theta\cos(\worb t - \xis)\Big] + \cos(\wrot t)\Big[\sin i\sin(\worb t)\cos\Theta - \cos i\sin\Theta\sin(\worb t - \xis )\Big]}{\sqrt{1 - (\cos i \cos \Theta + \sin i \sin\Theta \cos\xi_s)^2}\sqrt{1 - \sin^{2}\Theta\cos^{2}(\worb t - \xis)}}.
\end{equation}

In Schwartz et al. (2016) these results were arrived at using a series of rotations in an inertial frame. We rederive these results here using a more intuitive and simplified argument. This intuitive argument suffers from a minor sign ambiguity that we point out below; the energetic reader can confirm that we have selected the correct signs here by referring to the previous argument in Schwartz et al. (2016). 

The geometry is described, to within time-reversal, by three fixed vectors in an inertial frame: the orbital and spin angular momentum vectors, $\vec{\omega}_{\text{orb}} = \omega_{\text{orb}} \hat{z}$ and $\vec{\omega}_{\text{rot}}$, and the vector towards the observer $\hat{n}_o$, see Figure \ref{OrbGeometry}.  The angles between these three vectors are the planetary obliquity, $\Theta$, the orbital inclination, $i$, and the sub-observer longitude $\theta_o$. The angle between the plane spanned by $\hat{\omega}_{\text{orb}}$ and $\hat{n}_o$ and that spanned by $\hat{\omega}_{\text{orb}}$ and $\hat{\omega}_{\text{rot}}$ is the solstice phase $\xi_s$, as illustrated in inset (a) of the right panel of Figure \ref{OrbGeometry}. 
The equations (\ref{eq:c_thobs}-\ref{eq:s_phst}) can all be derived by applying the spherical law of cosines to this configuration and the variations of it involving $\hat{n}_s$ displayed in the right panel of Figure \ref{OrbGeometry}. Applying the spherical law of cosines to inset (a) gives equation \eqref{eq:c_thobs} and since $\theta_{o}$ is in the range $[0,\pi]$, this is enough to uniquely fix the sign of $\sin(\theta_o)$ to that of equation \eqref{eq:c_thobs}. Equations \eqref{eq:c_phobs} follow immediately from our assumptions on the initial condition for the sub-observer longitude $\phi_{o}(0) = 0$, hence $\phi_o = - \omega_{\text{rot}} t$. The equations for the sub-stellar colatitude (\ref{eq:c_thst}) follow from applying the spherical law of cosines to inset (d) and noting that $\xi = \worb t$. The last two equations for the sub-stellar longitude (\ref{eq:c_phst}-\ref{eq:s_phst}) are more complicated, but can still be reached in a few lines. 

A simple route to equation \eqref{eq:c_phst} is to rewrite
\begin{equation}
\cos(\phi_s) = \cos(\phi_o+\phi_s-\phi_o)= \cos(\phi_o) \cos(\phi_s-\phi_o)-\sin(\phi_o) \sin(\phi_s-\phi_o). 
\end{equation}
Now we use inset (b) of the right panel of Figure \ref{OrbGeometry} to find the relation 
\begin{equation}
\label{phdiff}
\cos(\phi_s - \phi_o) = \frac{\cos (\pi-w) - \cos \theta_o \cos \theta_s}{\sin \theta_o \sin \theta_s},
\end{equation}
again by the spherical law of cosines. Furthermore, from inset (c) we have 
\begin{equation}
\label{piwidth}
\cos (\pi-w) = \sin i \cos \xi = \sin i \cos( \worb t). 
\end{equation}
Plugging this result and equations (\ref{eq:c_thobs}) and (\ref{eq:c_thst}) into \eqref{phdiff} then gives
\begin{equation}
\label{c_longdiff}
\cos(\phi_s - \phi_o)  = \frac{\sin i \cos(\worb t)-\cos \theta_o \sin \Theta \cos(\worb t- \xi_s)}{\sqrt{1-(\cos i \cos \Theta + \sin i \sin\Theta \cos\xi_s)^2} \sqrt{1 - \sin^{2}\Theta\cos^{2}(\worb t - \xis)}}.
\end{equation}
Returning to equation \eqref{phdiff} we can also compute the sine of the difference in longitudes
\begin{equation}
\sin(\phi_s-\phi_o) = \sqrt{1- \left(\frac{\cos (\pi-w) - \cos \theta_o \cos \theta_s}{\sin \theta_o \sin \theta_s}\right)^2} = \frac{\sqrt{1-\cos^2 \theta_o - \cos^2 \theta_s - \cos^2 (\pi-w)+2 \cos \theta_o \cos \theta_s \cos(\pi -w)}}{\sin \theta_o \sin \theta_s}.
\end{equation}
Once again using \eqref{piwidth}, (\ref{eq:c_thobs}), and (\ref{eq:c_thst}) the numerator of this equation turns out to be a perfect square and we have 
\begin{equation}
\label{s_longdiff}
\sin(\phi_s-\phi_o) = \frac{\sin i\sin(\worb t)\cos\Theta - \cos i\sin\Theta\sin(\worb t - \xis )}{\sqrt{1-(\cos i \cos \Theta + \sin i \sin\Theta \cos\xi_s)^2} \sqrt{1 - \sin^{2}\Theta\cos^{2}(\worb t - \xis)}}.
\end{equation}
Inserting the results \eqref{c_longdiff} and \eqref{s_longdiff} into \eqref{phdiff} and using $\cos \phi_o = \cos(\wrot t)$ and $\sin \phi_o = -\sin(\wrot t)$ completes the derivation and gives \eqref{eq:c_phst}. The derivation for \eqref{eq:s_phst} proceeds along almost identical lines. The sign ambiguity inherent in this derivation is due to the choice of sign whenever square roots are used. The signs chosen in the formulas reproduce the argument of Schwartz et al. (2016), which is not subject to any such ambiguity. 

By making pairs of vectors parallel or orthogonal (setting the above angles to 0 or $\pi/2$), we arrive at 6 special cases: edge-on orbit ($i=\pi/2$), face-on orbit ($i=0$), pole-on observer ($\theta_o =0$), equatorial observer ($\theta_o =\pi/2$), zero-obliquity ($\Theta=0$), Cassini state ($\Theta=\pi/2$).  The substellar longitude is by far the ugliest expression in the general case, and the only way to significantly simplify it is by setting the planetary obliquity to zero.

One can combine various of the above cases (e.g., zero obliquity planet on edge-on orbit) to obtain yet simpler expressions. Finally, one can make various special cases and approximations regarding the two frequencies, $\omega_{\rm rot}$ and $\omega_{\rm orb}$, such as assuming that orbital motion is slow compared to the timescale of observations. These limits are easier to compute and can be used to explore the lightcurve signatures of $\delta$-maps. In the next few subsections we detail these special cases. We only report expressions that differ from the general case.

\subsection{Orbital Inclination}
For an edge-on orbit, $i=\pi/2$, the sub-observer co-latitude is $\cos\theta_o = \sin\Theta \cos\xi_s$, with concomitant changes in the sub-stellar longitude: 
\begin{equation}
	%\label{eq:c_phstip}
	\cos\phst = \frac{\cos(\wrot t)\Big\{\cos(\worb t) - \cos\thobs\sin\Theta\cos\left[ \worb t - \xis \right]\Big\} + \sin(\wrot t)\sin(\worb t)\cos\Theta}{\sin\theta_o\sqrt{1 - \sin^{2}\Theta\cos^{2}(\worb t - \xis)}},
\end{equation}
\begin{equation}
	%\label{eq:s_phstip}
	\sin\phst = \frac{-\sin(\wrot t)\Big\{\cos(\worb t) - \cos\thobs\sin\Theta\cos\left[ \worb t - \xis \right]\Big\} + \cos(\wrot t)\sin(\worb t)\cos\Theta\}}{\sin\theta_o\sqrt{1 - \sin^{2}\Theta\cos^{2}(\worb t - \xis)}}.
\end{equation}

For a face-on orbit, $i=0$, the sub-observer co-latitude is $\cos\theta_o = \cos\Theta$, and the sub-stellar latitude simplifies to:
\begin{equation}
	%\label{eq:c_phstiz}
	\cos\phst = \frac{-\cos(\wrot t) \cos\Theta\cos\left[ \worb t - \xis \right] - \sin(\wrot t)\sin \left[ \worb t - \xis \right]}{\sqrt{1 - \sin^{2}\Theta\cos^{2}\left[ \worb t - \xis \right]}},
\end{equation}
\begin{equation}
	%\label{eq:s_phstiz}
	\sin\phst = \frac{\sin(\wrot t)\cos\Theta\cos\left[ \worb t - \xis \right] - \cos(\wrot t)\sin \left[ \worb t - \xis \right]}{\sqrt{1 - \sin^{2}\Theta\cos^{2}\left[ \worb t - \xis \right]}}.
\end{equation}

\subsection{Subobserver Latitude}
The pole-on observer, $\theta_o=0$ or $\theta_o=\pi$, is a special case and cannot be obtained as a limit of the above expressions.  Nonetheless, expressions for the various angles were derived in Schwartz et al. (2016):
\emph{Case 1:} The sub-stellar point \emph{will not} pass over the poles during orbit:
\begin{equation}
	\cos\phst = \frac{\cos\wrot\cos\xi\cos\Theta + \sin\wrot\sin\xi}{\sqrt{1 - \sin^{2}\Theta \cos^{2}\left[ \xi - \xis \right]}}, \qquad \text{and} \qquad 	\sin\phst = \frac{-\sin\wrot\cos\xi\cos\Theta + \cos\wrot\sin\xi}{\sqrt{1 - \sin^{2}\Theta \cos^{2}\left[ \xi - \xis \right]}}.
\end{equation}

\emph{Case 2:} The sub-stellar point \emph{will} pass over the poles during orbit:
\begin{equation}
	\cos\phst = \frac{-\sin\wrot\sin\xi\cos\xis}{\sqrt{1 - \sin^{2}\Theta\cos^{2}\left[ \xi - \xis \right]}}, \qquad \text{and} \qquad 	\sin\phst = \frac{-\cos\wrot\sin\xi\cos\xis}{\sqrt{1 - \sin^{2}\Theta\cos^{2}\left[ \xi - \xis \right]}}.
\end{equation}

An equatorial observer, $\theta_o=\pi/2$, leads to the following expressions: 
\begin{equation}
	%\label{eq:c_phsteq}
	\cos\phst = \frac{\cos(\wrot t)\Big\{\sin i\cos(\worb t) \Big\} + \sin(\wrot t)\Big\{\sin i\sin(\worb t)\cos\Theta - \cos i\sin\Theta\sin(\worb t - \xis )\Big\}}{\sqrt{1 - \sin^{2}\Theta\cos^{2}(\worb t - \xis)}},
\end{equation}
\begin{equation}
	%\label{eq:s_phsteq}
	\sin\phst = \frac{-\sin(\wrot t)\Big\{\sin i\cos(\worb t) \Big\} + \cos(\wrot t)\Big\{\sin i\sin(\worb t)\cos\Theta - \cos i\sin\Theta\sin(\worb t - \xis )\Big\}}{\sqrt{1 - \sin^{2}\Theta\cos^{2}(\worb t - \xis)}}.
\end{equation}

\subsection{Planetary Obliquity}
If the planet has zero obliquity, $\Theta=0$, then $\theta_o=i$, $\theta_s=\pi/2$, and $\phi_s=(\worb-\wrot)t$. With these relations we have $\cos\thobs = \cos i$, $\cos\thst = 0$,
$\cos\phst = \cos[(\worb-\wrot)t]$, and $\sin\phst = \sin[(\worb-\wrot)t]$.

If the planet has a ninety degree obliquity, $\Theta=\pi/2$, then 
\begin{equation}
	%\label{eq:cs_thsteq}
	\cos\thobs = \sin i \sin\Theta \cos\xi_s, \qquad \text{and} \qquad
	\sin\thobs = \sqrt{1-\cos^2\thobs},
\end{equation}
\begin{equation}
	%\label{eq:cs_thsteq}
	\cos\thst = \cos(\worb t - \xis), \qquad \text{and} \qquad
	\sin\thst = \sin(\worb t - \xis ),
\end{equation}
\begin{equation}
	%\label{eq:c_phst}
	\cos\phst = \frac{\cos(\wrot t)\Big[\sin i\cos(\worb t) - \cos\thobs\cos(\worb t - \xis)\Big] - \sin(\wrot t)\cos i\sin(\worb t - \xis )}{\sqrt{1 - \sin^2 i \cos^2\xi_s}\sin(\worb t - \xis)},
\end{equation}
\begin{equation}
	%\label{eq:s_phst}
	\sin\phst = \frac{-\sin(\wrot t)\Big[\sin i\cos(\worb t) - \cos\thobs\cos(\worb t - \xis)\Big] - \cos(\wrot t)\cos i\sin(\worb t - \xis )}{\sqrt{1 - \sin^2 i \cos^2\xi_s}\sin(\worb t - \xis)}.
\end{equation}

\subsection{Rotational and Orbital Frequencies}
The synchronously rotating scenario is a subset of the zero-obliquity case. The orbital and rotational angular frequencies are equal, $\worb=\wrot$, so the the sub-stellar longitude is fixed, $\phi_s=0$.

If, instead, the planet's orbital period is much greater than its rotational period, then $\worb t \approx \xi_0$. We may then define the constant $\Xi \equiv \xi_0-\xi_s$, the sub-stellar co-latitude is constant, and all of the time-variability enters as $\wrot t$:
\begin{equation}
	%\label{eq:cs_thstfop}
	\cos\thst = \sin\Theta\cos\Xi, \qquad \text{and} \qquad
	\sin\thst = \sqrt{1 - \sin^{2}\Theta\cos^{2}\Xi},
\end{equation}
\begin{equation}
	%\label{eq:cs_phstfop}
	\cos\phst = \cos(\wrot t + \phi_{\rm hr}), \qquad \text{and} \qquad
	\sin\phst = \sin(\wrot t + \phi_{\rm hr}),
\end{equation}
where the hour angle of the observer is
\begin{equation}%\label{eq:a}
	\tan \phi_{\rm hr} = \frac{\cos i\sin\Theta\sin \Xi - \sin i\sin(\xi_0)\cos\Theta}{\sin i\cos(\xi_0) - \cos\thobs\sin\Theta\cos\Xi}.
\end{equation}

\section{Jacobian for Recursion Relation}
\label{Jacobian}

Using the expressions for the Euler angles and for the lune width in terms of the subobserver and substellar points, Eqs. \eqref{Eulers} and \eqref{width}, we are able to find the Jacobian between these variables:
\begin{equation}
\label{JacDet}
J \equiv \det \frac{\partial(\alpha, \beta, \gamma, w)}{\partial(\theta_s, \phi_s, \theta_o, \phi_o)}=\frac{ \sin \theta_o \sin \theta_s\left[ (1-\cos^2 \theta_s \cos^2 \theta_o)(\cos^2 \theta_s +\cos^2 \theta_o-2\cos^2 \theta_s \cos^2 \theta_o)+j(\theta_s, \phi_s, \theta_o, \phi_o)\right] }{\sin^2 w \left[\sin^2 w- \sin^2 \theta_s \sin^2 \theta_o \sin^2 (\phi_s-\phi_o)\right]^{3/2}},
\end{equation}
where the right hand side is only a function of $\{\theta_s, \phi_s, \theta_o, \phi_o\}$, but we have used $\sin w$ and $j(\theta_s, \phi_s, \theta_o, \phi_o)$ as convenient shorthands, use Eq. \eqref{width} to expand the first, and the second is defined by
\begin{equation}
\begin{aligned}
j(\theta_s, \phi_s, \theta_o, \phi_o) \equiv 2& \cos \theta_s \cos \theta_s \cos^3 (\phi_s-\phi_o) \sin^3 \theta_s \sin^3 \theta_o-\cos^2 (\phi_s-\phi_o) \sin^2 \theta_s \sin^2 \theta_o(\cos^2 \theta_s +\cos^2 \theta_o -6 \cos^2 \theta_s \cos^2 \theta_o)\\
&-2\cos \theta_s \cos \theta_o \cos (\phi_s-\phi_o) \sin \theta_s \sin \theta_o(1+\cos^2 \theta_s +\cos^2 \theta_o -3 \cos^2 \theta_s \cos^2 \theta_o).
\end{aligned}
\end{equation}
Except for when the subobserver or substellar points are at the North or South pole of the planet, the numerator only vanishes at isolated points. This shows that the Euler angles $\{ \alpha, \beta, \gamma \}$ and the lune width $w$ are in general independent variables and hence that we were justified in considering the partial derivative with respect to $w$ at fixed values of $\alpha, \beta,$ and $\gamma$. The recursion relation arrived at in Eq. \eqref{recurs} is rigorously correct whenever $J\neq 0$. 

\section{Explicit Examples of Analytic Harmonic Lightcurves}
\label{Explicit}
In this Appendix we list some explicit examples of analytic lightcurves. At the end of the Appendix we also illustrate the recursion of harmonic lightcurves, Eq. \eqref{recurs},  derived in Section \ref{Symm}.

\subsection{Tidally-locked and inclined analytic lightcurves}

For a tidally-locked and inclined orbit we can choose the planet coordinates so that $\theta_s = \pi/2$ and $\phi_s =0$. The vanishing obliquity allows simplification of the subobserver co-latitude
\begin{equation}
\label{incthobs}
\cos\thobs = \cos i \cos \Theta + \sin i \sin\Theta \cos\xi_s = \cos i,
\end{equation}
and this makes sense given our convention $i \in [0,\pi]$. Using Eqs. \eqref{Eulers} and \eqref{width} and taking care with their ranges, the Euler angles and lune width take on the simple values $\alpha = \pi/2$, $\gamma = \pi$,  
\begin{equation}
\cos \beta = -\frac{\sin \theta_o \sin \phi_o}{\sqrt{1- \sin^2 \theta_o \cos^2 \phi_o}}, \qquad \text{and} \qquad \cos w = -\sin \thobs \cos \phobs . 
\end{equation}
Putting these values in the general result Eq. \eqref{Light}, we tabulate several low-order harmonic light curves (we do not list lightcurves with $m<0$ since these can be found directly from the symmetry of Eq. \eqref{symm}):
\begin{equation}
\label{F10inc}
F_1^0(t) = \frac{1}{16}\sqrt{\frac{3}{\pi}}(1+\cos \phobs \sin \thobs)|\cos \thobs|,
\end{equation}
\begin{equation}
F_1^1(t) = -\frac{1}{16}\sqrt{\frac{3}{2\pi}}(1+e^{i \phobs} \sin \thobs)(1+\cos \phobs \sin \thobs) ,
\end{equation}
\begin{equation}
F_2^0(t) = -\frac{1}{6\pi \sqrt{5\pi}} \left[2\arccos(- \cos \phobs \sin \thobs)\cos \phobs \sin \thobs +(2-3 \cos^2 \thobs)\sqrt{1-\cos^2 \phobs \sin^2 \thobs} \right],
\end{equation}
\begin{equation}
F_2^1(t) = -\frac{1}{\pi \sqrt{30 \pi}}\left[\arccos(- \cos \phobs \sin \thobs) +e^{i \phobs} \sin \thobs \sqrt{1-\cos^2 \phobs \sin^2 \thobs} \right] | \cos \thobs |,
\end{equation}
\begin{equation}
F_2^2(t) = \frac{1}{2 \pi \sqrt{30 \pi}}\left[2 e^{i \phobs} \arccos(- \cos \phobs \sin \thobs) \sin \thobs+(1+e^{2i \phobs} \sin^2 \thobs) \sqrt{1-\cos^2 \phobs \sin^2 \thobs} \right]  .
\end{equation}
Equation \eqref{incthobs} allows us to identify $\thobs = i$, and so the time-dependence of these light curves can be made completely explicit by substituting $\phobs = -\wrot t$, as argued in Appendix \ref{geometry}.

\subsection{A completely general analytic harmonic lightcurve and an illustrative recursion}
Comparison with CFH13 alone does not display the full power of our main result because that paper never dealt with non-zero obliquities. Here we illustrate the power of our main result with a completely general analytic lightcurve and discuss a subtlety. The subtlety has do with limiting a general lightcurve to the special tidally locked geometry or to a second special geometry discussed below. We have already seen, in Section \ref{checking}, that these  cases exhibit subtleties like gimbal lock and special values of the $\arctan(x)$ function. If you are interested in a particular geometry where the Euler angles of Eq. \eqref{Eulers} may take special values, it is wise to take a look at these formulas by hand at first. 

The first non-Lambertian lightcurve with an arbitrary viewing geometry and expressed in terms of the Euler angles and lune width is
\begin{equation}
F_1^0 = -\frac{1}{4}\sqrt{\frac{3}{\pi}} \cos \left( \frac{w}{2}+\gamma \right) \sin^3 \frac{w}{2} \sin \beta.
\end{equation}
This can be rewritten in terms of the subobserver and substellar points
\begin{equation}
\label{F10}
F_1^{0} = \frac{1}{16}\sqrt{\frac{3}{\pi}}\,   \,  (\cos \theta_o+  \cos \theta_s)[1+\cos \thst \cos \thobs+\sin \thst \sin \thobs \cos(\phst-\phobs)],
\end{equation}
 and substituting the time dependence of the angles from Appendix \ref{geometry} this reduces to the remarkably simple and unexpected form
\begin{equation}
\begin{aligned}
\label{explicit}
F_1^0(t) = \frac{1}{16}\sqrt{\frac{3}{\pi}}\, \, [\cos i \cos \Theta + \sin i \sin\Theta \cos\xi_s + \sin \Theta \cos(\worb t - \xi_s)] [1+\sin i \cos(\worb t)].
\end{aligned}
\end{equation}
This equation makes the dependence of $F_1^0(t)$ on the orbital geometry completely explicit. 

However, the attentive reader may balk at Eqs. \eqref{F10} and \eqref{explicit} because in the zero obliquity limit, where $\thst = \pi/2$ and $\phi_s =0$, they disagree with Eq. \eqref{F10inc}; no $|\cos \thobs |$ appears in \eqref{F10} in this limit. The reason behind this observation highlights the care needed in taking the limit. Technically the Euler angle $\gamma$ is given by 
\begin{equation}
\gamma = \text{atan2}\,(\cos \thst, -\frac{\cos \thobs + \cos \thst \cos w}{\sin w})
\end{equation}
where $\text{atan2}\,(y,x) = \arctan (y/x)$ returns an angle in the full $(-\pi,\pi]$ range given the components $y$ and $x$. The arctangent function has a branch cut, which can be taken along the negative $x$-axis, and hence is discontinuous whenever $y=0$ and one varies $x$ from positive to negative values. In the present example, when $\thst = \pi/2$ the first argument of $\text{atan2}\,(y,x)$ vanishes and $\gamma$ discontinuously jumps from $\pi$ for a northern observer to $0$ for a southern observer (note that $\sin w$ is always positive). This change is exactly what accounts for the absolute values in Eq. \eqref{F10inc} and is completely absent at any other substellar co-latitude. 

If a lightcurve depends on $\alpha$, e.g. $F_1^1$, then the analogous branch cut discontinuity needs to be taken into account. The full definition of $\alpha$ is 
\begin{equation}
\alpha = \text{atan2}\,(\cos \thobs \sin \thst \cos \phst-\cos \thst \sin \thobs \cos \phobs, \cos \thst \sin \thobs \sin \phobs - \cos \thobs \sin \thst \sin \phst).
\end{equation}
The branch cut discontinuity is again on the line where the first argument vanishes, that is, when
\begin{equation}
\cos \thobs \sin \thst \cos \phst-\cos \thst \sin \thobs \cos \phobs = 0.
\end{equation}
Geometrically this condition corresponds to the lune pole lying either on the planet's prime meridian or on its 180th meridian. When this is the case there is a discontinuity in $\alpha$ whenever the lune pole passes through the planet's North or South poles (i.e. when $\cos \thst \sin \thobs \sin \phobs - \cos \thobs \sin \thst \sin \phst$ goes from positive to negative). This is the second special geometry mentioned at the beginning of this section.  

Finally we illustrate the general recursion of Eq. \eqref{recurs}. Using the main result, Eq. \eqref{Light}, we first compute $F_1^1$ 
\begin{equation}
F_{1}^1 = -\frac{1}{8} \sqrt{\frac{3}{2 \pi}} e^{i \alpha} \sin^2 \frac{w}{2} \left[ \cos \left( \frac{w}{2}- \beta+\gamma \right) +\cos \left( \frac{w}{2}+\beta+\gamma \right) +2 i \sin \left( \frac{w}{2}+\gamma \right) \right],
\end{equation}
and then check 
\begin{equation}
\left(\frac{-2i}{\sin \beta} \frac{\partial}{\partial \gamma} -2 \cot \beta\right) F_{1}^{1} = -\frac{1}{4}\sqrt{\frac{6}{\pi}} \cos \left( \frac{w}{2}+\gamma \right) \sin^3 \frac{w}{2} \sin \beta e^{i \alpha}= \sqrt{2} F_{1}^{0} e^{i \alpha}
\end{equation}
which is exactly Eq. \eqref{recurs} with $l=m=1$.

\end{document}